\documentclass[fleqn,usenatbib]{mnras}

\usepackage{newtxtext,newtxmath}
\usepackage[T1]{fontenc}

\DeclareRobustCommand{\VAN}[3]{#2}
\let\VANthebibliography\thebibliography
\def\thebibliography{\DeclareRobustCommand{\VAN}[3]{##3}\VANthebibliography}

\usepackage{graphicx}	% Including figure files
\usepackage{amsmath}	% Advanced maths commands
\newcommand{\nh}{{\sc NewHorizon}}

\title[Dwarf structure in data and theory]{2D light distributions of dwarf galaxies -- key tests of the implementation of physical processes in simulations
}

\author[A. E. Watkins et al.]{
A. E. Watkins,$^{1}$\thanks{E-mail: a.watkins@herts.ac.uk (AEW)}
G. Martin,$^{2}$
S. Kaviraj,$^{1}$
C. Collins,$^{3}$
Y. Dubois,$^{4}$
K. Kraljic,$^{5}$
C. Pichon,$^{6}$
S. K. Yi,$^{7}$
\\
$^{1}$Centre for Astrophysics Research, University of Hertfordshire, College Lane, Hatfield AL10 9AB, UK\\
$^{2}$School of Physics and Astronomy, University of Nottingham, University Park, Nottingham NG7 2RD, UK\\
$^{3}$Astrophysics Research Institute, Liverpool John Moores University, IC2 Building, Liverpool Science Park, 146 Brownlow Hill, Liverpool L3 5RF, UK\\
$^{4}$Institut d'Astrophysique de Paris, UMR 7095, CNRS, Sorbonne Universit\'e, 98 bis boulevard Arago, 75014 Paris, France\\
$^{5}$Universit\'{e} de Strasbourg, CNRS, Observatoire astronomique de Strasbourg, UMR 7550, F-67000 Strasbourg, France\\
$^{6}$Institut d’Astrophysique de Paris, Sorbonne Universit\'{e}, CNRS, UMR 7095, 98 bis bd Arago, 75014 Paris, France\\
$^{7}$Department of Astronomy and Yonsei University Observatory, Yonsei University, Seoul 03722, Korea
\\
}

\pubyear{2024}

\begin{document}
\label{firstpage}
\pagerange{\pageref{firstpage}--\pageref{lastpage}}
\maketitle

\begin{abstract}
Cosmological simulations provide much of the theoretical framework within which we interpret extragalactic observations.  However, even if a given simulation reproduces the integrated properties of galaxies well, it may not reproduce the detailed structures of individual galaxies.  Comparisons between the 2D light distributions of simulated and observed galaxies -- particularly in the dwarf regime, where key processes like tidal perturbations and baryonic feedback most strongly influence galaxy structure -- thus provide an additional valuable test of the simulation's efficacy.  We compare scaling relations derived from mock observations of simulated galaxies, drawn from the two largest halos in the high-resolution \nh{} cosmological simulation, with galaxies in the Fornax cluster.  While Fornax is significantly more massive than either group, it is the lowest-mass cluster in the local Universe, and contains a well-studied population of spatially resolved dwarfs, hence serves as a useful benchmark.  Per unit stellar mass, \nh{} dwarfs are systematically larger in half-light radius, much fainter in surface brightness, and bluer in colour than their Fornax counterparts, albeit with similar light profile shapes.  We discuss potential reasons for these discrepancies, including environmental effects, baryonic feedback, resolution, or couplings of these factors.  As observations of dwarfs outside of the local Universe become more plentiful through on-going or up-coming surveys such as Euclid and LSST, 2D comparisons such as these, where properties are measured in the same way across both simulations and observations, can place strong constraints on processes that alter the spatial distribution of baryons in galaxies.
\end{abstract}

\begin{keywords}
galaxies: dwarf -- galaxies: evolution -- galaxies: fundamental parameters -- galaxies: structure -- galaxies: clusters: general -- galaxies: clusters: individual: Fornax
\end{keywords}

%%%%%%%%%%%%%%%%%%%%%%%%%%%%%%%%%%%%%%%%%%%%%%%%%%

%%%%%%%%%%%%%%%%% BODY OF PAPER %%%%%%%%%%%%%%%%%%

\section{Introduction}

The distribution of baryons within galaxies is the result of a long and complex interplay between many internal and external forces.  For example, once cold gas forms into stars, the most massive of those stars eject energy and metals back into the gas during their lifespans (through stellar winds) and in their explosive death knells \citep[e.g.][]{capriotti01, oppenheimer06, hopkins12, dale14, kobayashi20}, altering the means by which subsequent stellar populations can form.  Likewise, collisions with other galaxies can rearrange the stellar or dark matter phase-space distributions of the involved systems \citep[e.g.][]{toomre72, moore98, mastropietro05, binney08, mendezabreu12, elichemoral18, martin18}, permanently altering the galaxies' subsequent dynamical evolution.

Tracking the relative influences of these disparate phenomena on any individual galaxy's evolution is difficult in the real Universe, and so we rely on large-scale simulations to construct narratives matching observations \citep[e.g.][]{springel05, dubois14, vogelsberger14, schaye15, kaviraj17, springel18}.  Necessarily, however, these simulations are constructed using physical assumptions derived from observations, so the two regimes---theoretical and observational---inform and build off of each other.  While the origin and evolution of large-scale structure, such as galaxy clustering, is well-constrained in our current cosmological paradigm of $\Lambda$CDM \citep[e.g.][]{guo11}, fully realizing the detailed small-scale evolution of galaxies remains a work in progress.

Proposed resolutions to some small-scale problems, such as the "missing satellites" problem \citep{klypin99, moore99} or the "too big to fail" problem \citep{boylankolchin11}, rely on a confluence of factors, including the relationship between baryonic and dark matter, the completeness limits of existing all-sky surveys, and the impact of environment or baryonic feedback on gas retention and star formation \citep[e.g.][]{guo11, sawala16, kim2018, jackson21}.  The latter factor is critical, as without some kind of preventative feedback, stars condense in the centers of massive galaxies and in low-mass galaxies far too quickly \citep[e.g.][]{white78, cole91, white91, beckmann17}.

The choice of feedback model is not a trivial one, however, as different models can produce degenerate results \citep[e.g.][and references therein]{wright24}.  Some degeneracy is, of course, expected: the broad impact of feedback is to push against runaway gravitational collapse, and so serves as a means for systems to self-regulate.  However, in modern simulations, the feedback is often calibrated to reproduce a narrow set of fundamental scaling relations, such as the stellar mass function \citep[e.g.][]{dayal14, mccarthy17} or the cosmic star formation history \citep[e.g.][]{schaye10}.  A narrow scope can neglect important details, as even the stellar mass function can vary with environment \citep[e.g.][]{yang09, wetzel13, monterodorta21}, for example.

Dwarf galaxies can provide a significant additional calibration for simulations. They have, by definition, low masses, and therefore shallow potential wells, which makes their star formation histories particularly sensitive to key processes like feedback \citep[e.g.][]{lacey91, read05, hopkins12, onorbe15, davis22}. Additionally, gas loss via ram pressure stripping (RPS) is very efficient in dwarfs \citep{boselli08, toloba15, venhola19, boselli22}, and tidal interactions tend to be more effective at disrupting them or rearranging mass within them than in higher mass galaxies \citep{moore98, mastropietro05, koch12, montes21, jackson21b, jang24}.  Dwarf galaxies are also the most common kind of galaxy by number in all environments \citep{driver94, blanton05, mcnaughtroberts14, kaviraj17}, meaning that they can provide a large, robust sample from which to draw conclusions about the physics driving galaxy evolution.

A comparison between real and simulated dwarfs is thus an effective means of calibrating a simulation at a granular level.  Beyond the stellar mass function, galaxies follow scaling relations between stellar mass, size, surface brightness, light concentration, and other structural properties \citep[e.g.][]{okamura84, bershady00, conselice03, shen03, saintonge08, holwerda14, vanderwel14, paulinoafonso19, trujillo20}.  Following the evolution of these relations through cosmic time and across environments probes how they must arise from the same self-regulated evolution of baryonic mass that drives the stellar mass function \citep[e.g.][]{trujillo06, szomoru12, huertascompany13, taylor16, hamadouche22}.  The 2D distribution of baryons is thus a more stringent test of a simulation's efficacy.  Reproducing outliers from these relations \citep[for example, dwarfs with large size and low surface brightness;][]{sandage84, impey88} may be even more critical, as such galaxies could be the most sensitive probes of feedback and gravitational physics available \citep[e.g.][]{dicintio17, jackson21}.

Ideally, an investigation into dwarf galaxies would use samples spanning many environments, from voids to massive clusters.  However, obtaining accurate distances to dwarfs is a time-intensive process \citep[e.g.][]{zaritsky22}, and so surveys targeting dwarf galaxies are often centred around massive systems with which the dwarfs can be associated spatially \citep[e.g.][]{geha17, venhola18, trujillo21}.  In such cases, the impacts of baryonic feedback on dwarf galaxy structure cannot be easily isolated from those of environment \citep[e.g.][]{watkins23}.  That said, clusters and massive groups---representing the most extreme density environments---tend to produce extreme impacts on their dwarf populations \citep{sandage84, ferguson90, lisker06, janz21, romerogomez24}, and of course contain large numbers of galaxies of all stellar mass, including dwarfs.  Thus, for a comparison between simulations and observations, clusters and massive groups do offer the following: abundant populations of heavily processed dwarfs, complete down to low stellar mass.

We thus compare the structural scaling relations of dwarf galaxies found in the two most massive halos in the \nh{} simulation \citep{dubois21} with the dwarf sample unveiled in the Fornax Cluster by \citet{venhola18}, the Fornax Deep Survey \citep[FDS;][]{peletier20} Dwarf galaxy Catalogue (FDSDC).  Comparisons with the FDSDC benefit from its abundant supply of dwarfs in a dense environment nearby enough that detailed structural decompositions are possible \citep[e.g.][]{su21}.  We begin, in Sec.~\ref{sec:data}, with a brief overview of both the \nh{} simulation and the FDS. We then describe the methods by which we construct and analyse synthetic observations from \nh{} in Sec.~\ref{sec:methods}.  We compare the simulated dwarf and FDS dwarf scaling relations in Sec.~\ref{sec:results}.  We explore potential reasons for the differences we uncover between FDS and \nh{} dwarfs in Sec.~\ref{sec:discussion}. 
 Finally, we summarize the paper in Sec.~\ref{sec:summary}.

%%%%%%%%%%%%%%%%%%%%%%%%%%%%%%%%%%%%%%%%%%%%%%%%%
\section{Simulated data and observations}\label{sec:data}

\subsection{New Horizon Simulation}\label{ssec:nh}

\nh{} is a zoom-in simulation of a spherical region with a diameter of 20~Mpc within its parent, \textsc{horizon-agn} \citep{dubois14, kaviraj17}, large enough for a 4096$^{3}$ resolution (cells; in terms of dark matter mass, M$_{\rm DM}=1.2\times10^{6}$ M$_{\odot}$).  \nh{} thus uses the same cosmological model as \textsc{horizon-agn}, with parameters compatible with WMAP-7 \citep{komatsu11}: matter density $\Omega_{\rm m}=0.272$, dark energy density $\Omega_{\Lambda}=0.728$, amplitude of the matter power spectrum $\sigma_{8}=0.81$, baryonic density $\Omega_{\rm b}=0.045$, Hubble constant $H_{0}=0.74$ km s$^{-1}$ Mpc$^{-1}$, and scalar spectral index $n_{\rm s}=0.967$.

\nh{} is run with the adaptive mesh refinement {\sc RAMSES} code \citep{teyssier02}, resulting in a maximum spatial resolution of 34~pc within a contiguous $(16$~Mpc$)^{3}$ volume.  Combined with the very high stellar mass resolution ($1.3\times10^{4}$
M$_{\odot}$), this makes the simulation very useful for a study of dwarf galaxies.  One limitation of our study, however, is that of environment, as the volume sampled is not large enough to include any clusters with Fornax-like halo masses.  Instead, we draw our dwarf population from {\sc NewHorizon}'s two largest groups, with halo masses of $6\times10^{12}$ M$_{\odot}$ and $7\times10^{12}$ M$_{\odot}$.  These are approaching the mass of Fornax \citep[$7\pm2\times10^{13}$ M$_{\odot}$;][]{drinkwater01}, but are about an order of magnitude smaller.  We consider the implications of this difference throughout our study.

Star formation in \nh{} occurs above a gas density threshold of $n_{0}=10$~cm$^{-3}$, following a Schmidt relation \citep{schmidt59} wherein the star formation rate (SFR) is proportional to the gas density and inversely proportional to the local free-fall time, with a variable efficiency \citep{padoan11, federrath12, kimm17}.  Likewise, the turbulent Mach number is determined by the local 3D instantaneous velocity dispersion.  SFRs (and, consequently, feedback) can occur at higher gas densities and thus achieve higher values than in its parent simulation, depending on local conditions.

Feedback itself is limited to Type II supernovae (SNe) and active galactic nuclei (AGN).  The latter is unimportant in the simulated dwarf regime \citep[the requisite black holes tend not to grow in \nh{} dwarf galaxies;][]{jackson21}, and hence will not be considered here.  The former releases $10^{51}$ ergs of kinetic energy per explosion, assuming a lower limit of $6$ M$_{\odot}$ for stars which can explode.  The frequency of such stars assumes each star particle ($10^{4}$ M$_{\odot}$) is composed of a simple stellar population following a Chabrier IMF with a mass range of $0.1 < $M$/$M$_{\odot} \leq 150$ \citep{chabrier05}.  Momentum is transferred via the mechanical SN feedback scheme of \citet{kimm14, kimm15}.

\nh{} reproduces many observed galaxy scaling relations to good accuracy, particularly at the high-mass scale.  This includes, for example, the galaxy mass function, the cosmic star formation rate and stellar density, the Kennicutt-Schmidt relation \citep{schmidt59, kennicutt98}, the Tully-Fisher relation \citep{tully77}, the massive black hole-to-galaxy mass relation \citep{dubois21}, and many others using integrated quantities.

\subsection{Fornax Deep Survey}\label{ssec:fds}

We compare the simulated dwarfs from \nh{} with observed dwarfs found in the Fornax Cluster via the Fornax Deep Survey, which combines data from the VLT Survey Telescope (VST) Early-Type GAlaxy Survey (VEGAS, PIs: M. Capaccioli and E. Iodice) and the FOrnax Cluster Ultra-deep Survey \citep[FOCUS, PI: R. F. Peletier;][]{capaccioli15}.  Both of these were conducted with the VST's square-degree OmegaCAM \citep{kuijken02} camera at the European Southern Observatory, targeting both the Fornax Cluster itself (26 deg$^{2}$ centred on the brightest cluster galaxy NGC~1399 in $g^{\prime}$, $r^{\prime}$, and $i^{\prime}$, and 21 deg$^{2}$ in $u^{\prime}$) and its associated sub-group Fornax A.  In the three deepest photometric bands ($g^{\prime}$, $r^{\prime}$, and $i^{\prime}$), the FDS reaches limiting surface brightnesses $>30$ mag arcsec$^{-2}$ \citep[3$\sigma$, 10\arcsec$\times$10\arcsec;][]{venhola18}.  \citet{iodice16} and \citet{venhola18} provide a detailed account of the survey's observation strategy and data reduction procedure.  For all of our derived quantities, we assume a distance to the Fornax Cluster of 20~Mpc \citep{blakeslee09}.

For our study, we used the photometric parameters derived for the FDS Dwarf galaxy Catalogue \citep[FDSDC;][]{venhola18} by \citet{su21} and later \citet{watkins23} using a combination of radial profiles and S\'{e}rsic$+$point spread function (PSF) decompositions.  The catalogue contains such parameters for 564 dwarf galaxies in Fornax with stellar masses as low as $10^{5}$ M$_{\odot}$ \citep[with said stellar masses estimated from $g^{\prime}-i^{\prime}$ and $r^{\prime}-i^{\prime}$ colours using the calibrations from][]{taylor11}.

For our comparison with \nh{}, we limit the sample to galaxies with stellar masses $> 10^{7}$ M$_{\odot}$ (corresponding to $\sim1000$ \nh{} star particles), as any simulated dwarfs with masses lower than this will not be very well resolved.  While the \citet{su21} catalogue does not include the 265 additional dwarfs identified later by \citet{venhola22}, only seven additional galaxies with $\log($M$_{*}/$M$_{\odot}) > 7$ were found among the latter, meaning the resolution limits of \nh{} preclude a realistic comparison with most of these newly uncovered dwarfs.  The \citet{su21} catalogue still contains 250 galaxies with masses between $7 < \log($M$_{*}/$M$_{\odot}) < 9.5$, significantly more than either \nh{} group (likely due to Fornax's larger halo mass), hence our study is limited (in terms of number statistics) mostly by the simulated catalogue.

%%%%%%%%%%%%%%%%%%%%%%%%%%%%%%%%%%%%%%%%%%%%%%%%%
\section{Methods}\label{sec:methods}

Here we briefly summarize the methods we employed to create synthetic observations from the \nh{} simulation, and to derive structural parameters from those synthetic observations.  We note that we include all of the massive ($\log($M$_{*}/$M$_{\odot} > 9.5$) FDS and \nh{} group galaxies alongside our analysis of dwarfs, as \nh{} uses the massive galaxy population as their typical benchmark for observational comparisons.

\subsection{Synthetic image generation}\label{ssec:synth}

\begin{figure*}
    \centering
    \includegraphics[scale=1.0]{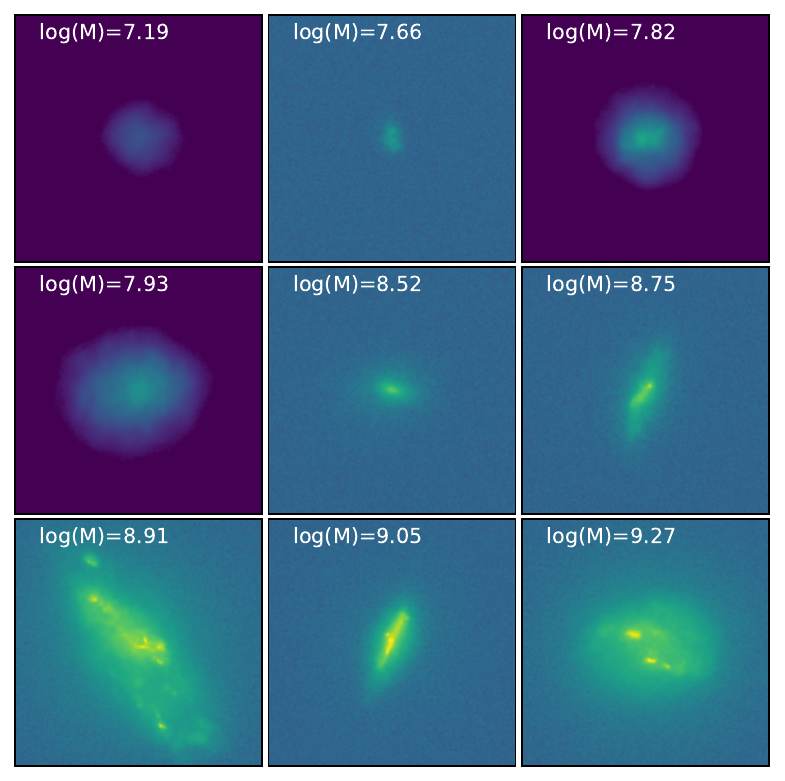}
    \caption{Nine example stamps made of synthetic \nh{} dwarfs, at the final \nh{} timestep ($z=0.17$).  All images are scaled such that the width is 10~kpc, and the surface brightness spans $\mu_{r^{\prime}}=29.5$--$21$ mag arcsec$^{-2}$.  We mark the galaxy stellar masses in the upper left of each panel, in units of $\log($M$_{\odot})$.  Panels with dark backgrounds (top and middle left, top right) show examples of galaxies} too faint to be detected using the FDS sky background (around half the sample; see text), whose background we adjusted to be arbitrarily low-noise.
    \label{fig:stamps}
\end{figure*}

To compare the photometric properties of simulated dwarfs with those of real dwarfs, we constructed synthetic observations of all dwarf (and massive) galaxies found within the two largest halos in the \nh{} simulation.  We did this following the process outlined by \citet{martin22}, with parameters meant to replicate the observations made for the FDS.  While we provide a brief summary here, for a full description of the image-generation process, we refer the reader to that paper.  Additional details will be elaborated on by Martin et al. (in prep.).

We first extracted all star particles within a 20~kpc radius centred on each galaxy with stellar mass above M$_{*}=10^{10}$M$_{\odot}$, or a 10~kpc radius otherwise.  As this does not exclude any particles within that volume, we are not reliant on any specific galaxy identification methodology, other than the values of stellar mass and centroid; we discuss one consequence of this in Sec.~\ref{ssec:photometry}.  As each particle is described by an age and a metallicity, we derived particle-by-particle SEDs via interpolation of a grid of simple stellar population (SSP) models from \citet{bruzual03}, assuming a \citet{chabrier03} initial mass function (IMF), which is also what was used to calculate the integrated stellar masses we compare against.  We used a screen dust model along the line of sight to each particle to simulate attenuation of these SEDs, where the dust column density is given by
\begin{equation}\label{eq:dust}
    N_{\rm cell} = \rho Z \Delta r \times {\rm DTM}.
\end{equation}
Here, $\rho$ is the gas density in the adaptive mesh refinement cell, $Z$ is the gas-phase metallicity, $\Delta r$ is the line of sight cell length, and DTM is the dust-to-metal ratio.  For the latter, we used two values: above $\log($M$_{*}/$M$_{\odot}) \geq 10$, we adopted the value from \citet{martin22} of $\textrm{DTM}=0.4$ \citep{draine07}; below this, we used $\textrm{DTM}=0.1$.  The latter we chose based on the relation between DTM and metallicity shown in Fig.~4 of \citet{li19}, using the median galaxy mass in our simulated sample of $\log($M$_{*}/$M$_{\odot}) = 8$ and the mass-metallicity relation for dwarfs shown in Fig.~6 of \citet{buzzo24}.

We then produced dust-attenuated SEDs as:
\begin{equation}\label{eq:extinction}
    I(\lambda)_{\rm attenuated} = I(\lambda)e^{-\kappa(\lambda)N}\,,
\end{equation}
where $\kappa(\lambda)$ is the dust opacity, and where $N$ is the total column density in front of each star particle, summed along the line of sight.  We derived dust opacities from the \citet{weingartner01} extinction curves, using the Milky Way (MW) $R_{V}=3.1$ curve for galaxy masses above $\log($M$_{*}/$M$_{\odot}) \geq 10$, and the Small Magellanic Cloud (SMC) bar extinction curve below this.  Magnitudes and luminosities were derived by convolving these attenuated SEDs (corrected for redshift) with the $g^{\prime}$ and $r^{\prime}$ filter transmission curves.  We show examples of nine \nh{} galaxy stamps created through this process in Fig.~\ref{fig:stamps}.

We note that the choice of DTM and extinction curve has little impact on the magnitudes and surface brightnesses we derived.  Varying the DTM from 0.01 to 1.0, and using either MW or SMC extinction curves results in variation in total magnitudes of order $0.005$~mag and in central surface brightnesses of order $0.01$~mag arcsec$^{-2}$ in both $g^{\prime}$ and $r^{\prime}$.  This is likely because these galaxies are located in massive groups and so are relatively gas-poor compared to the field population.  Likewise, we expect using full radiative transfer rather than a screen dust model would not impact our conclusions \citep[see e.g. Fig.~5 of][]{kaviraj17}.  We find that the use of a \citet{salpeter55} IMF primarily affects the bluest galaxies, decreasing their magnitudes by $\sim 1$ in either band, and their surface brightnesses by $\sim0.6$ mag arcsec$^{-2}$ \citep[echoing][]{martin22}.  Red galaxies are mostly unaltered.

For low-density regions within the galaxies, we applied the same adaptive smoothing algorithm used by \citet{martin22}\footnote{\url{https://github.com/garrethmartin/smooth3d}}, which follows a method similar to that of {\sc ADAPTIVEBOX}, employed by \citet{merritt20}.  In summary, we split each star particle in these low-density regions into 500 equal flux particles with normally distributed positions, centred at the original particle position with a standard deviation equal to the particle's fifth nearest neighbour distance, then generated the image by summing the sub-particle fluxes along a chosen axis.  The latter step employs a 2D grid with elements 0.2\arcsec$\times$0.2\arcsec, following the FDS pixel scale.  We then created image stamps of these flux-scaled, smoothed particle distributions, along the simulation's $xy$ axis (an arbitrary choice, meaning the galaxy orientations are random).  While the particle data is drawn from the simulation's final time step of $z=0.17$, we scale the size and flux of each galaxy to $z=0.0047$, which yields a distance modulus of 31.51 \citep[echoing][]{blakeslee09} using the WMAP-7 \citep{komatsu11} cosmological parameters.  At this scale, one pixel is $\sim 20$~pc, which is smaller than the highest-resolution cells in the \nh{} simulation; however, we found by creating a subset of more distant stamps that this has little impact on our derived photometric quantities.

Once stamps of the simulated galaxies are created in this way, we apply image characteristics to them similar to the FDS coadds used by \citet{venhola18} to construct the FDSDC.  These coadded images have a pixel scale of 0.2 arcsec px$^{-1}$, with all flux scaled such that the AB magnitude zeropoints are 0.0 in all photometric bands.  To mimic the survey resolution, we convolved the simulated galaxies in our images with the analytic PSF models published by \citet{venhola18}.  These models are comprised of Gaussian$+$Moffat profile cores (their Eq.~5), with wings modelled as declining exponential functions (their Eq.~6, extending to 160\arcsec \ radius).  Given that the PSF is variable from field to field, we used the PSF parameters for Field 11 \citep[Table~A2 of][]{venhola18}, in the centre of Fornax, to construct our model FDS PSFs for each photometric band.  As both \citet{venhola18} and \citet{su21} used only the $g^{\prime}$ and $r^{\prime}$ bands for their photometric analysis, we constructed only these images for the \nh{} sample.

Also following \citet{martin22}, we added synthetic sky-subtracted backgrounds to all mock exposures using the coadded images' limiting surface brightnesses converted to a per-pixel $1\sigma$ variance \citep[their Eq.~3, derived from Appendix A of][]{roman20}.  These are $\mu_{{\rm lim},\,g^{\prime}} = 28.4$ mag arcsec$^{-2}$ and $\mu_{{\rm lim},\,r^{\prime}} = 27.8$ mag arcsec$^{-2}$ \citep[$1\sigma$, $1$\arcsec$\times1$\arcsec;][]{venhola18}.  However, forty-three of the \nh{} galaxies were too faint to be detectable at this surface brightness, so for these we used an arbitrary background of $\mu_{{\rm lim}, \lambda}=32$ mag arcsec$^{-2}$ to avoid throwing out nearly half the sample.

\subsection{Photometry and decompositions}\label{ssec:photometry}

Having produced mock FDS observations of the \nh{} galaxies, for a fair comparison, we needed to derive their photometric properties in the same ways in which they were derived for the real FDS dwarfs.  The photometric techniques employed by \citet{su21} for the FDS galaxies follow those developed for the \emph{Spitzer} Survey of Stellar Structure in Galaxies \citep[S$^{4}$G;][]{sheth10} by \citet{munozmateos15} and \citet{salo15}.  Again, we provide only a brief summary of these methods here, pointing out any differences in technique we employ for this work; a full description can be found in the papers cited above.

From every synthetic \nh{} exposure, we derived radial surface brightness profiles and curves of growth, from which we derived integrated properties such as the effective radius.  We measured these profiles using the Astropy-affiliated package {\sc Photutils's} \citep[v.1.11.0;][]{bradley24} isophote-fitting algorithm {\sc ellipse}, itself based on the {\sc IRAF} routine of the same name \citep{jedrzejewski87, busko96}.

We first fit the galaxies using free parameters to establish their characteristic isophotal shapes.  For each galaxy, we estimated the galaxy centres initially using the {\sc Photutils} center-of-mass centroiding algorithm within a $20\times20$~px box at the image centre.  We then ran {\sc ellipse} multiple times per galaxy, cycling through a handful of starting isophotal parameter (semi-major axis length, position angle, and ellipticity) combinations until each fit proceeded successfully.  Unlike in real images, which contain masked interloping sources, correlated noise and other similar features, we found that successful fits were robust to these starting parameters, so no fine-tuning was necessary.  If a fit proceeded with a set of starting parameters, it would proceed identically on a successful fit using a different set of starting parameters.

With free-parameter fits made, we again measured the radial profiles using fixed isophote shapes derived from the outer regions of each galaxy \citep[see][]{munozmateos15}.  We chose these by examining each galaxy's free-parameter radial position angle (PA) and ellipticity ($\epsilon$) profiles, hand-selecting radius ranges where these parameters were roughly constant, and taking the median PA, $\epsilon$, and centre coordinates measured within those limits as the characteristic isophotal parameters for the galaxy.  This was often difficult, as unlike the the Fornax galaxies, almost all of these synthetic galaxies showed highly variable isophote shapes with radius.  To assist with this, we also created deprojected images of each galaxy using these parameters by transforming each image's Cartesian coordinates into elliptical coordinates, accepting the parameters as valid only if these deprojected galaxy images appeared roughly circular.  We derived all photometric quantities (such as $R_{\rm eff}$) from either the surface brightness profiles or curves of growth measured from the standard projection images using the fixed isophotal parameters derived in this way.

Normally, prior to measuring magnitudes, sizes, and so on via the resulting curves of growth, a local background estimate is made and subtracted by measuring flux in apertures near the galaxies but visibly dominated by background noise.  As we injected backgrounds with zero mean flux into our synthetic images, initially we deemed such a correction unnecessary.  However, we found through examination of sky-less stamps that many dwarfs are embedded in a near-uniform, diffuse medium at very low surface brightness ($\mu_{r}>32$ mag~arcsec$^{-2}$), which contributes non-negligible flux to the lowest-mass dwarfs when its contribution is summed over large concentric annuli.  This originated from how we built our stamps; as stated in Sec.~\ref{ssec:synth}, we did not exclude any particles within a 10~kpc (20~kpc for massive galaxies) radius volume around each galaxy.  We therefore apply background corrections to our photometry, measured in boxes placed at the stamp edges, to correct for this.  The necessity of such corrections implies a systematic uncertainty on our measurements which is not present when estimating simulated galaxy sizes via the particle distributions or structure finder directly, but which is always present when using similar techniques on images of real galaxies.

\citet{su21} also performed S\'{e}rsic profile decompositions using the two-dimensional fitting algorithm {\sc Galfit} \citep{peng02, peng10}.  We did this as well, first deriving single-component decompositions for all galaxies, then multi-component decompositions for each galaxy which seemed to merit additional components based on the appearance (from visual inspection) of the residuals from the best-fit single-component models.  If the stamp contained an obvious companion galaxy, we fit both simultaneously.  We used these to estimate S\'{e}rsic indices and central surface brightnesses of the galaxies, and to compare the frequency with which multi-component fits are required for \nh{} dwarfs compared to Fornax dwarfs.

{\sc Galfit} requires two templates for optimal fitting results: a normalized PSF model, and a sigma image \citep[used to weight each pixel in the fit; see][]{peng02}.  For the former, we used the models with which we convolved the simulated galaxies described in the previous section.  For the latter, we derived weights for each pixel using the following equation:
\begin{equation}\label{eq:sigmaimage}
    \sigma(x,y) = [({\rm gain} \cdot (I(x,y) + S))^{2} + {\rm RON}^{2}]^{1/2}\,,
\end{equation}
where gain is the camera gain in electrons per analogue-to-digital unit (e$^{-}$ ADU$^{-1}$), $I(x,y)$ is the synthetic galaxy's intensity at pixel $(x,y)$ in ADU, $S$ is a constant sky brightness in ADU, and RON is the readout noise in e$^{-}$.  We used the mean of the gains found in the headers of the FDSDC stamps we had available as the gains per band, and we set the RON to 4.4 e$^{-}$, the mean of 19 CCDs in VST's OmegaCam\footnote{\url{https://www.eso.org/sci/facilities/develop/detectors/optdet/docs/papers/omegacam_poster.pdf}}.  For $S$, we used the mean sky brightnesses for $g^{\prime}$ and $r^{\prime}$ published in Table~2 of \citet{yoachim16}, converted to ADU.

We found using a handful of stamps that {\sc Galfit} produces very similar fits when using its own auto-generated sigma images compared to those we supply, assuming the proper information is present in the stamp headers.  Supplying our own sigma images therefore serves mostly to speed up the fitting procedure.  Likely this is due to the idealized nature of the synthetic exposures, which lack artefacts, masked pixels, and have perfectly Gaussian noise backgrounds by construction.  We also found that the {\sc Galfit} results, if the fits succeeded, were insensitive to the initial parameters supplied, other than the number of iterations required to reach a minimum in the $\chi^{2}$ between the model and the image.

Five of the \nh{} galaxies appeared to be extremely diffuse and without any discernible structure.  These have anomalously faint magnitudes for their stellar masses, and are not well characterized by S\'{e}rsic profiles.  Indeed, their light profiles are not monotonically declining, meaning derived parameters such as total magnitude (and thus, half-light radius) are ambiguous at best.  We have verified that the fraction of low-resolution dark matter particles is zero among all group galaxies we study, including these \citep[see also][]{jang24}, thus they do not appear to be dwarf galaxies altered by resolution effects.  Likely these were simply identified as local peaks within extremely underdense regions of the galaxy groups by the structure finder (possibly tidal features or intragroup light); hence, we remove these from our sample.  Likewise, some dwarfs were embedded within the isophotes of companion galaxies; for these, we use the \textsc{Galfit} magnitudes and sizes, rather than the curve of growth measures, as the latter is contaminated by the companion's light.  Finally, the \nh{} sample contains ten extremely compact objects reminiscent of ultra-compact dwarfs \citep[UCDs;][]{drinkwater00}, which are not present in the FDSDC (although they are present in the Fornax Cluster itself).  As our interest here is normal dwarf galaxies, we identify these as $>5\sigma$ outliers from the half-light radius--stellar mass relation and remove them from the sample as well.  \nh{} UCDs are discussed in detail by \citet{jang24}.

%%%%%%%%%%%%%%%%%%%%%%%%%%%%%%%%%%%%%%%%%%%%%%%%%
\section{Results}\label{sec:results}

\begin{figure*}
    \centering
    \includegraphics[scale=0.675]{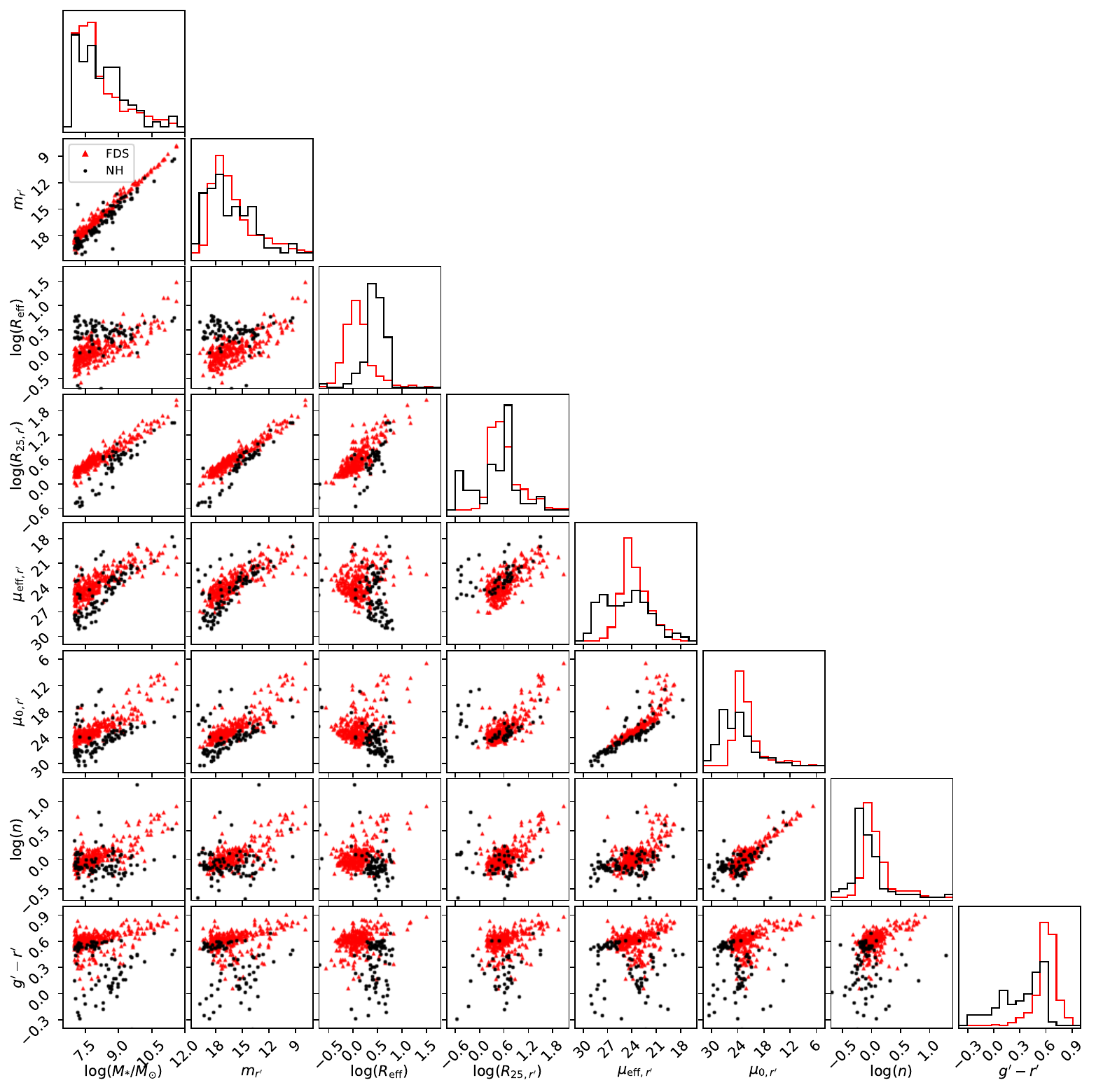}
    \caption{A corner plot, showing the correlations between eight different variables.  From the left to right along the bottom axis, these are: stellar mass (in solar masses), integrated $r^{\prime}$-band magnitude, effective radius (kpc, estimated along the major axis), the $\mu_{r^{\prime}}=25$ mag arcsec$^{-2}$ isophotal radius (kpc, with surface brightness corrected for inclination using the axial ratio), mean surface brightness within the effective radius, central surface brightness, S\'{e}rsic index, and integrated $g^{\prime}-r^{\prime}$ color.  In all panels, black dots or histograms indicate \nh{} synthetic galaxies, while red triangles or histograms indicate FDS galaxies.  All sizes and other photometric quantities here were derived in the $r^{\prime}$ photometric band.}
    \label{fig:relations}
\end{figure*}

Fig.~\ref{fig:relations} shows a variety of correlations between derived galaxy properties in the form of a corner plot.  In each panel, black dots denote \nh{} galaxies, while red triangles denote Fornax Cluster galaxies.  Histograms of each distribution along the diagonal are similarly colour-coded.  Descriptions of the parameters plotted are provided in the figure caption.  Here we relate the most significant similarities and differences between the simulated and observed cluster galaxies.

At stellar masses below $\log($M$_{*}/$M$_{\odot}) \lesssim 9$, we find a slight ($\sim$1~mag) offset in the mass-magnitude relation, where agreement is much better above this threshold.  Use of a Salpeter IMF does not affect this offset at the low-mass end, but does create a similar offset at higher stellar masses.  This implies that \nh{} dwarfs are under-luminous compared to their Fornax counterparts.  Partly this may be an artifact of our derivation of SEDs using the \citet{bruzual03} SSP models, which span a limited number of metallicities.

Dwarfs show an offset in the mass-size relation between \nh{} and FDS.  For galaxies below $\log($M$_{*}/$M$_{\odot}) \lesssim 9.5$, $R_{\rm eff}$ is systematically too high among \nh{} galaxies.  Here, the mass-size relation is roughly horizontal at $R_{\rm eff} \sim 3$~kpc, while the FDS dwarfs show a distinct positive correlation between $R_{\rm eff}$ and stellar mass despite the somewhat large scatter.  We note this size offset can also be seen in Fig.~4 of \citep{jang24}.

\nh{} dwarfs also have much smaller 25 mag arcsec$^{-2}$ isophotal radii ($R_{25,\,r^{\prime}}$) than the FDS dwarfs.  This is due to their lower surface brightnesses: \nh{} shows mostly parallel $\mu$--$\log($M$_{*})$ relations to FDS, offset to fainter values (although the matches appear better at higher stellar mass).  Here, the offset is much larger than that with magnitude, at 3--4 mag arcsec$^{-2}$.  This is larger for central surface brightness $\mu_{0}$, which can be seen as a slight offset between FDS and \nh{} in the $\mu_{0,\,r^{\prime}}$--$\mu_{{\rm eff},\,r^{\prime}}$ relation, suggesting that the offset is not uniform with radius.  Correcting dwarf surface brightness by increasing $m_{r^{\prime}}$ to match the FDS $m_{r^{\prime}}$--$\log($M$_{*})$ relation only improves the match slightly, implying that it is the larger values of $R_{\rm eff}$ causing this difference.

The distribution of S\'{e}rsic indices ($n$) among \nh{} and FDS dwarfs is fairly similar, scattering mostly around $n=1$, though \nh{} dwarfs appear to trend slightly below their FDS counterparts.  FDS dwarfs show a slight positive trend in $n$ with stellar mass \citep[for a discussion of this, see][]{watkins23}, which is absent among the \nh{} dwarfs.  In general, however, \nh{} and FDS show a fairly good match in terms of dwarf light profile shapes (despite, as stated above, the normalizations of those profiles being different).

\nh{} dwarfs show a wider range of integrated colour ($g^{\prime}-r^{\prime}$) than their FDS counterparts.  Most of the \nh{} dwarfs in these groups appear quenched below $\log($M$_{*}/$M$_{\odot}) \lesssim 8.5$, with uniformly red colours.  These still lie at the blue end of the FDS colour distribution.  While this may imply that \nh{} dwarfs are lower in metallicity, it may be another artefact of our use of the \citet{bruzual03} models to generate the mock SEDs.  Above this mass threshold, most \nh{} galaxies are significantly bluer than those in Fornax, implying higher SFRs.  This may be a result of the differences in halo mass: these groups are both a factor of $\sim 10$ lower in virial mass than the Fornax Cluster, implying that RPS is not as efficient here.

Finally, we find that $20$ per cent of the \nh{} dwarf galaxies require multiple components when fitting {\sc Galfit} models, always in the form of central light concentrations.  From \citet{su21}, 25 per cent of the Fornax dwarfs have bright, compact nuclei, a comparable fraction.

In summary, \nh{} dwarfs are fainter, more radially extended, lower in surface brightness, and bluer in colour than FDS dwarfs at the same stellar mass, despite having similar roughly exponential light profile shapes.  We explore possible reasons for these differences in the next section.

%%%%%%%%%%%%%%%%%%%%%%%%%%%%%%%%%%%%%%%%%%%%%%%%%
\section{Discussion}\label{sec:discussion}

\nh{} dwarfs in the two largest groups found in the simulation are larger and more diffuse than Fornax Cluster dwarfs of the same stellar mass.  Here, we explore the most likely reasons for these differences.  We split these into two broad categories: environment (including RPS and tidal interactions), and internal processes (feedback).  Throughout, we also consider the joint impact of these two factors, as well as the issue of the simulation's resolution.

\subsection{Environmental influence}\label{ssec:environment}

\begin{figure*}
    \centering
    \includegraphics[scale=1.0]{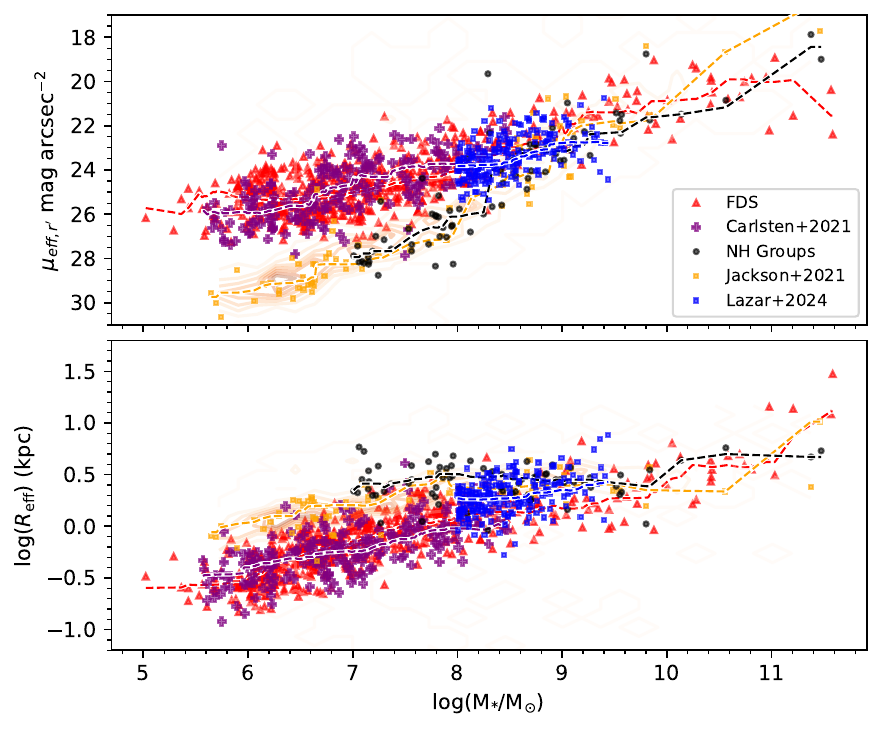}
    \caption{Comparing effective surface brightness and effective radius vs. stellar mass relations from a variety of different studies.  As Fig.~\ref{fig:relations}, red triangles show the FDS relation and black points show our NH group dwarf values.  Purple crosses show values for Local Volume dwarf satellites from \citet{carlsten21}.  Orange squares denote measurements by \citet{jackson21} also of galaxies in the two largest \nh{} groups, but made using the unattenuated intensity-weighted second moments of the particle distributions.  Orange contours underlying these points outline the distributions for all \nh{} galaxies from \citet{jackson21}.  Blue squares are from \citet{lazar24a}, for $z<0.08$ dwarfs from the COSMOS2020 catalogue.}
    \label{fig:mueffcomp}
\end{figure*}

Perturbations from neighbouring galaxies, including fly-by interactions and merger events (in which two or more galaxies coalesce), can provide kinematic kicks to their existing gaseous, stellar, and dark matter components.  The impact on gas is different than on the collisionless stellar and dark matter components, as gas is able to radiatively cool and condense at high enough densities \citep[e.g.][]{lucy77, hernquist89}.  The impact of tidal perturbations on gas thus, broadly, leans toward compression and cooling, leading to temporary enhancements in star formation.  Most often this enhancement occurs in the galaxy core in conjunction with SFR suppression in the outskirts, as gas loses angular momentum in response to torques \citep[e.g.,][]{keel85, mihos94, ellison08, moreno15, daikuhara24}.  One might naively expect that interaction-induced star formation should ultimately serve to enhance the central densities of galaxies, decreasing their $R_{\rm eff}$ and increasing $\mu_{\rm eff}$ or $\mu_{0}$.  However, this study demonstrates that the \nh{} dwarfs show the opposite trend compared to the observed dwarfs.

The impact of tides on collisionless particles (stars and dark matter) depends on the parameters of the interaction.   For a given two-body system, the duration over which the tidal force is present determines the extent to which stellar orbits alter in response.  For example, whether an encounter is pro- or retrograde affects the length and thickness of any resulting tidal arms \citep{toomre72}.  Stars in a dwarf or globular cluster plunging through a more massive galaxy can also increase their velocity dispersion if they do not have enough time to adapt to the rapidly changing gravitational potential \citep[i.e., their dynamical times are long compared to the gravitational impulse, a process known as tidal shocking;][]{ostriker72}.  Stars within $1R_{\rm eff}$, with short dynamical times, are less impacted by tides.  In the most extreme such cases, the most loosely bound stars will become unbound \citep[e.g.,][]{penarrubia08}, although on non-radial orbits this may require several pericentric passages given the stability provided by their dark matter halo.  Broadly, such interactions tend to result in a $<1$ per cent loss in stellar luminosity even in the dwarf regime \citep[e.g.][]{mihos04, martin22}.

Stellar velocity dispersion can also increase in response to a sudden dramatic mass-loss, such as the loss of gas during RPS 
\citep[e.g.][]{hammer24}.  This is simply a gravitational response to the decreased potential well depth, as, assuming no external tidal perturbation, the stars in systems undergoing RPS retain their original orbital velocities.  However, in a simulation by \citet{boselli08}, it was found that the impact of RPS on cluster dwarf structures at long wavelengths ($H$-band) was negligible, while at shorter wavelengths, RPS ultimately resulted in decreased surface brightness (due to truncated star formation) and effective radius (due to an enhanced SFR ratio between the core and the disk).  The ultimate impact of RPS on dwarf structure thus depends on whether or not the gas is removed predominantly via the stripping itself (thereby altering the dwarf's gravitational potential), or via enhanced star formation induced by the ram pressure (thereby merely rearranging the existing mass within said potential).

Additionally, environment impacts the amount of fuel available for star formation, especially at high redshift.  Denser regions promote initially faster gas-infall, resulting in higher initial SFRs.  If this star formation is then later truncated by interactions or RPS within said dense environment, these initial conditions can impact the evolutionary end-state of the galaxies \citep[e.g.][]{martin19, jackson21}.  In this way, environment and baryonic feedback can end up coupled.

That said, observations indicate that the scaling relations we investigate here may not be strongly correlated with environment.  \citet{habas20} and \citet{poulain21} showed that their sample of dwarfs from the MATLAS survey \citep{duc15}, preferentially surrounding massive early type galaxies, falls along the same magnitude-$R_{\rm eff}$ and magnitude-surface brightness relations as cluster, Local Volume, and Local Group dwarfs.  We show similar results in Fig.~\ref{fig:mueffcomp}, which compares the correlations between $\mu_{\rm eff}$ and $R_{\rm eff}$ with stellar mass across four different studies.  As in Fig.~\ref{fig:relations}, we show the FDS relation as red triangles and our \nh{} group dwarf relation as black points.  Purple crosses denote measurements from \citet{carlsten21}, for a sample of Local Volume satellite dwarfs.  Orange squares denote measurements for \nh{} galaxies in the two most massive groups by \citet{jackson21}, who estimated surface brightnesses using the unattenuated intensity-weighted second moments of the particle distribution rather than via 2D analysis of mock images (as we did).  Orange contours underlying these points showcase the entire \nh{} galaxy population from \citet{jackson21}.  Blue squares are values from the $z<0.08$ field dwarf galaxy sample from \citet{lazar24b,lazar24a} derived from the COSMOS2020 catalogue \citep{weaver22}.  The coloured dashed lines represent the running medians of the corresponding datasets.  For better comparison, we include the full range of stellar masses from \citet{venhola18} for the FDS sample in this figure.

Despite the different techniques used for estimating $\mu_{{\rm eff},\,r^{\prime}}$, our \nh{} $\mu_{{\rm eff},\,r^{\prime}}$--$\log($M$_{*}/$M$_{\odot})$ relation aligns well with that from \citet{jackson21}.  Both of these lie systematically below the relations derived from observations at low stellar mass, all of which agree well between studies despite the differences in sampled populations.  This implies that the structural differences we find between \nh{} and observed dwarfs are not limited to those found in the massive halos.  While tidal interactions and RPS can influence dwarf structure, it may be that the competition between the impacts of tidal disturbance or shocking (increasing stellar velocity dispersion, thus lowering surface brightness) and enhancements in SFR (increasing central stellar mass and central surface brightness) dampens the impact of these factors on the dwarf scaling relations we examine here.

We do see differences between our estimates of $R_{\rm eff}$ and those of \citet{jackson21}.  While the two distributions match somewhat well above $\log($M$_{*}/$M$_{\odot}) > 8$, our size estimates consistently lie above those of \citet{jackson21} for $\log($M$_{*}/$M$_{\odot}) \leq 8$, with the disparity increasing at lower stellar mass. The 2D curve of growth method thus appears to over-estimate $R_{\rm eff}$ for low stellar mass dwarfs compared to the intensity-weighted second-moment estimates.  As we mentioned in Sec.~\ref{sec:methods}, many of our galaxy stamps show extremely faint, diffuse backgrounds extending to the edges of the stamp boundaries.  While this light is extremely faint, it contributes significantly to the total fluxes of these low surface brightness dwarfs when summed up within large annulus apertures.  While we attempted to subtract its influence from our curves of growth, if the regions we sampled are unrepresentative of the diffuse background cospatial with the galaxies, these corrections may not have been sufficient.  Thus, this is a danger of employing standard observational photometric techniques on synthetic galaxy stamps generated in this manner.  Interestingly, a recent work investigating dwarf mass-size relations using the FIREbox simulation \citep{feldmann23}, which has a similar resolution to \nh{} ($\sim6\times10^{4}$M$_{\odot}$ stellar mass particles), by \citet{mercado25} shows a distribution in $R_{\rm eff}$--$\log(M_{*})$ very similar to that of \citet{jackson21}, implying the dwarfs are uniformly over-extended in that study as well.

Detailed examination indicates that some environmental differences in these scaling relations do exist.  For example, the $R_{\rm eff}$--$\log($M$_{*})$ relation for the field dwarf sample from \citet{lazar24a} is offset by $\sim\textrm{+}0.5$~kpc from the FDS relation \citep[though they may simply be extending off of the known late-type galaxy relation; e.g.][]{munozmateos15, watkins22}.  However, the most prominent difference is not between environments but between observed and simulated galaxy relations.  Indeed, the close agreement between our $\mu_{\rm eff}$--$\log($M$_{*})$ relation, for \nh{} dwarfs residing in the two most massive group halos, and that of \citet{jackson21}, which includes all \nh{} dwarfs, suggests that these simulated dwarfs are seemingly as unperturbed by environment as their observed counterparts.  Also, despite the uncertainty imposed by the stamps' diffuse backgrounds, both estimates of $R_{\rm eff}$ show that \nh{} dwarfs are larger everywhere.

This is worth investigating in more detail, but for the purposes of this paper, it implies that our choice to use Fornax Cluster dwarfs does not strongly influence the comparison in this parameter space.  Environmental influence appears to be subtle, with the most notable influence occurring in the faint outskirts of galaxies \citep[e.g.][]{chamba24}.

\subsection{Baryonic feedback}\label{ssec:feedback}

If environment has little direct impact on dwarf structure, baryonic feedback may provide most of the required energy.  Feedback is necessary to counterbalance gas cooling and inflows in the early Universe \citep{white78, cole91}, and, on more local scales, is needed to reproduce the flat, "cored" dark matter profiles often observed through kinematics in real galaxies \citep[e.g.,][]{pontzen12,jackson24} and to resolve the "missing satellite" and "too big to fail" problems \citep[e.g.,][]{zolotov12, wetzel16, garrisonkimmel19}.  The broad impact of feedback is thus toward radial diffusion of stellar and dark matter profiles, leading to lower surface densities.  However, precisely how this feedback impacts dwarf structure can depend on the specific sub-grid physical models employed by the simulation \citep[e.g.,][]{crain15, munshi19}, including how radiative cooling and the SFR couple to the simulation's resolution \citep[e.g.,][]{crain15, benincasa16, ludlow20}.  Isolating these effects is not a trivial task, but we can use results from previous work in the dwarf regime to aid our interpretation.

The baryonic feedback in \nh{} is limited to two kinds: Type II SNe and AGN \citep{dubois21}, both of which can transfer substantial energy and momentum into the surrounding gas (although AGN feedback in the dwarf regime here is limited due to a lack of supermassive black hole growth).  The ultimate result of either mechanism is a regulation between the gas inflow rate and the SFR \citep[e.g.][]{hopkins18}.

The details of this regulation matter, however.  In the early stages of growth, for example, when galaxies are composed of very few particles, energy and momentum can be distributed uniformly, as smaller structures like chimneys or bubbles remain unresolved \citep[e.g.][]{hopkins18}.  If the gas is not replenished, the resulting galaxies will be superheated, pressure-supported systems---at later epochs, then, most of the mass buildup would be via accretion of higher angular momentum gas, resulting in more extended systems \citep[indeed, with aggressive enough such feedback, this can even occur in Milky Way-mass galaxies; e.g.][]{roskar14}.  Additionally, momentum imparted by SN feedback can compress existing gas at lower densities, resulting in extended star formation \citep[e.g.][]{kimm15, jackson21, martinalvarez23}, which can dilute the stellar profile further, leading to the formation of low surface brightness systems \citep[e.g.][]{martin19, jackson21}.

\nh{} lacks gentler forms of feedback which can regulate the impact of SNe.  In real galaxies, young stellar clusters are surrounded by HII regions, which result from the ionizing radiation those young stars produce.  Because this emerges from the photopheres of massive stars, this feedback sets in before SNe.  It has a similar impact as SN feedback insofar as it prevents runaway collapse, but it does so without significant mass-loading \citep[e.g.][]{rosdahl15, emerick18, agertz20}.  Radiation, and stellar winds, thus can reduce SFRs \citep[e.g.][]{emerick18}, resulting in fewer SNe and thus less momentum transfer per unit time.

Of course, \nh{} uses various prescriptions to take such effects into account.  For example, the SN rate is enhanced over what is predicted by the chosen IMF, to simulate the influence of gas rarification via the collected influence of clustered SNe \citep[following][]{kim17, gentry19}.  Additionally, per-SN momentum is increased to take into account the radiative feedback from OB stars, following \citet{geen15}.  While such assumptions are theoretically well-motivated, enhancing either the SN-rate or the momentum transferred from them likely also serves to enhance the mass-loading.

Directionality of feedback can also influence how this momentum is distributed.  Real HII regions often show patchy morphology \citep[e.g.][]{hannon19}, implying that the corridors through which SN momentum propagates in this way is more constrained \citep[sometimes very much so, e.g.][]{kim23}.  While such effects may average out in more massive galaxies, these effects may be important in shaping galaxies in the dwarf regime.

In the absence of environment causing the discrepencies described above, baryonic feedback---likely coupled with resolution effects---is likely to be the primary cause behind the increased diffuseness of the \nh{} dwarfs.  However, understanding the precise origins of this diffuseness requires identifying the timestep at which the simulated and observed galaxies begin to diverge in size and surface brightness.  This, in turn, requires a comparison of real galaxy structural scaling relations across cosmic time, potentially to very high redshift, which is beyond the scope of the current study.  High-resolution space-based instruments such as JWST \citep{gardner06} should prove useful for making such a comparison in the future.

%%%%%%%%%%%%%%%%%%%%%%%%%%%%%%%%%%%%%%%%%%%%%%%%%
\section{Summary}\label{sec:summary}

Using synthetic images and existing derived parameters, we compared dwarf galaxy scaling relations between those found in the two largest groups in the \nh{} simulation with those measured from the Fornax Deep Survey (FDS).  Despite the differences in mass between these groups and Fornax, the latter contains one of the best available catalogues of spatially resolved dwarf galaxy parameters in a dense environment, hence is useful as a comparison.  We employed the same techniques to estimate these parameters from \nh{} as were employed in the FDS, thus removing as much methodological uncertainty as possible from the comparison.

While the \nh{} dwarfs have similar S\'{e}rsic indices and have a similar fraction containing central light concentrations as their Fornax counterparts, they differ in most other parameters.  \nh{} dwarfs are (per unit stellar mass) less luminous, more extended in half-light radius, bluer in colour, and fainter in surface brightness (central and effective) than dwarfs found in the Fornax Cluster.  The starkest contrast is in surface brightness, with \nh{} dwarfs being $3$--$4$ mag arcsec$^{-2}$ fainter than their Fornax counterparts.  This difference remains even after correcting for the offset in luminosity, implying that the starlight in \nh{} dwarfs is more radially extended than it should be.

Comparison with dwarfs from other observational studies shows that observed dwarfs vary only subtly with environment in mass-size (half-light radius) and mass-surface brightness relations.  Likewise, comparing our results in \nh{} with those of a previous study \citep{jackson21} shows that the simulation--observation offset is present across all of \nh{}, not just the two largest groups (and appears to be independent of how size is measured from the particle data).  Together, this implies that the diffuseness of \nh{} dwarfs is not primarily a result of environment.

Instead, baryonic feedback, likely coupled with resolution effects (particularly at high redshift), seems the more prominent cause.  The precise impact of feedback in a simulation is difficult to determine, however, given the aforementioned coupling with resolution, as well as the evolution over time as the gas, stellar, and dark matter mass in any individual galaxy grows through accretion.  Without an array of observed dwarf galaxy scaling relations to compare with out to high redshift, we cannot determine the precise conditions under which the mass-size offset sets in.  Identifying this mechanism using future observations should help refine feedback prescriptions in simulations to come, resulting in more realistic simulated low-mass galaxies.

Ultimately, we have demonstrated that a simulation which reproduces observed integrated galaxy properties may not necessarily reproduce the detailed structures of those same galaxies.  These 2D properties can thus provide important additional constraints on feedback prescriptions (including how these interact with the simulation's resolution), which can significantly impact their predictive power.  Given their sensitivity to such prescriptions, dwarfs provide an ideal laboratory for the processes that shape the spatial structure of galaxy components, and so the large dwarf samples expected to be observed in and outside the local Universe in large surveys such as Euclid \citep{laureijs11} or the Legacy Survey of Space and Time \citep[LSST;][]{ivezic19} will help broaden the scope of these comparisons to a wide variety of local conditions and redshifts.

%%%%%%%%%%%%%%%%%%%%%%%%%%%%%%%%%%%%%%%%%%%%%%%%%
\section*{Acknowledgements}

We thank the anonymous referee for their thorough and helpful report, which improved the clarity and quality of this manuscript.  SK and AEW acknowledge support from the STFC [grant number ST/X001318/1]. SK also acknowledges a Senior Research Fellowship from Worcester College Oxford. GM acknowledges support from the STFC under grant ST/X000982/1.  S.K.Y. acknowledges support from the Korean National Research Foundation (2020R1A2C3003769, RS-2022-NR070872).  This work made use of Astropy:\footnote{http://www.astropy.org} a community-developed core Python package and an ecosystem of tools and resources for astronomy \citep{astropy:2013, astropy:2018, astropy:2022}.  This research made use of Photutils, an Astropy package for detection and photometry of astronomical sources \citep[v.1.11.0;][]{bradley24}.  This work was granted access to the HPC resources of CINES under the allocations 2013047012, 2014047012, 2015047012, c2016047637, A0020407637 made by GENCI and KSC-2017-G2-0003 by KISTI, and as  a "Grand  Challenge"  project  granted  by  GENCI  on  the  AMD-Rome  extension of the Joliot Curie supercomputer at TGCC, and under the the allocation 2019-A0070402192 made by GENCI. This research is part of the  Segal (ANR-1919-CE31-0017, \url{http://secular-evolution.org}) and Horizon-UK projects. This work has made use of the Infinity cluster on which the simulation was post-processed, hosted by the Institut d’Astrophysique de Paris. We warmly thank S. Rouberol for running it smoothly.

%%%%%%%%%%%%%%%%%%%%%%%%%%%%%%%%%%%%%%%%%%%%%%%%%%
\section*{Data Availability}

The simulation data analysed in this paper were provided by the \nh{} collaboration. The data will be shared on request to the corresponding author, with the permission of the \nh{} collaboration or may be requested from \url{https://new.horizon-simulation.org/data.html}.  The observational data underlying this article are available at the Centre de Donn\'{e}es Astronomique de Strasbourg (CDS), at \url{ 10.26093/cds/vizier.36470100} and \url{https://cdsarc.cds.unistra.fr/viz-bin/cat/J/A+A/660/A69}, and at the European Southern Observatory Science Archive Facility, at \url{https://archive.eso.org/cms.html}.  Isophotal radii for FDS galaxies are published as supplementary data by \citet{watkins23}.
 
%The inclusion of a Data Availability Statement is a requirement for articles published in MNRAS. Data Availability Statements provide a standardised format for readers to understand the availability of data underlying the research results described in the article. The statement may refer to original data generated in the course of the study or to third-party data analysed in the article. The statement should describe and provide means of access, where possible, by linking to the data or providing the required accession numbers for the relevant databases or DOIs.



\bibliographystyle{mnras}
\bibliography{references} % if your bibtex file is called example.bib

%\begin{thebibliography}{99}
%\end{thebibliography}

%%%%%%%%%%%%%%%%%%%%%%%%%%%%%%%%%%%%%%%%%%%%%%%%%%

%%%%%%%%%%%%%%%%% APPENDICES %%%%%%%%%%%%%%%%%%%%%

%%%%%%%%%%%%%%%%%%%%%%%%%%%%%%%%%%%%%%%%%%%%%%%%%%

% Don't change these lines
\bsp	% typesetting comment
\label{lastpage}
\end{document}